\documentclass[
]{ceurart}

\usepackage{listings}

\begin{document}

\copyrightyear{2025}
\copyrightclause{Copyright for this paper by its authors.
  Use permitted under Creative Commons License Attribution 4.0
  International (CC BY 4.0).}

\conference{D-SAIL Workshop - Transformative Curriculum Design: Digitalisation, Sustainability, and AI Literacy for 21st Century Learning, July 22, 2025, Palermo, Italy}

\title{Our Coding Adventure: Using LLMs to Personalise the Narrative of a Tangible Programming Robot for Preschoolers}


\author{Martin Ruskov}[
orcid=0000-0001-5337-0636,
email=martin.ruskov@unimi.it,
]
\address{Department of Languages, Literatures, Cultures and Mediations, University of Milan, Piazza Sant'Alessandro 1, 20123 Milan, Italy}


\begin{abstract}
Finding balanced ways to employ Large Language Models (LLMs) in education is a challenge due to inherent risks of poor understanding of the technology and of a susceptible audience. This is particularly so with younger children, who are known to have difficulties with pervasive screen time.
Working with a tangible programming robot called Cubetto, we propose an approach to benefit from the capabilities of LLMs by employing such models in the preparation of personalised storytelling, necessary for preschool children to get accustomed to the practice of commanding the robot.
We engage in action research to develop an early version of a formalised process to rapidly prototype game stories for Cubetto. Our approach has both reproducible results, 
because it employs open weight models, and is model-agnostic, because we test it with 5 different LLMs.
We document on one hand the process, the used materials and prompts, and on the other the learning experience and outcomes. We deem the generation successful for the intended purposes of using the results as a teacher aid. Testing the models on 4 different task scenarios, we encounter issues of consistency and hallucinations and document the corresponding evaluation process and attempts (some successful and some not) to overcome these issues.
Importantly, the process does not expose children to LLMs directly. Rather, the technology is used to help teachers easily develop personalised narratives on children's preferred topics.
We believe our method is adequate for preschool classes and we are planning to further experiment in real-world educational settings.
\end{abstract}


\maketitle

\section{Introduction}

Employing Large Language Models (LLMs) in education is not a straightforward task~\cite{harvey_dont_2025}. It is particularly challenging with younger children, who are already known to have difficulties with pervasive screen time~\cite{wang_status_2023}.
At the same time, exactly in the early stages of education, children are very different and less used to a standardised educational process. They are less familiar to the abstractions necessary to develop early-stage competences in science, technology, engineering and mathematics (STEM). To illustrate this, already a non-trivial question is ``Why should I program a robot to go somewhere, when I can put it there myself?'' and they are not much appreciative of the need to develop computational thinking~\cite{macrides_programming_2022}.

Our contribution is in the context of Cubetto, a physical programming platform: a robot travelling on wheels and a control board with physically-insertable command blocks~\cite{solid_beginning_2017}. Whereas the creator-provided teaching materials accompanying Cubetto aim to teach how to use Cubetto, including physical and creative activities for children~\cite{solid_coding_2017}, they leave it to teachers to elaborate the extensive repetition that would allow children to get accustomed to the robot.
We set out from the premise that such getting used to Cubetto and its control is an important prerequisite for its adoption. Thus, we propose an approach to support pre-school teachers in developing the educational storytelling necessary to engage children with the learning activity. This is inspired by the narrative success of AI Dungeon, an interactive platform that engages users with ChatGPT to collaboratively create narrative experiences~\cite{hua_playing_2020}.
Using LLMs for such a learning activity that is not overly challenging allows for easier discernment of the desired pedagogical features.

\section{Background}

Computational thinking has been recognised as a key capability that children need to develop in the 21st century and programming skills are a cornerstone to it. Particularly from the 1980s, the practice of turtle programming has stuck as one of the most common onboarding approaches for younger children~\cite{grover_computational_2013,mcnerney_turtles_2004}. A number of platforms have attempted to make this accessible for even younger audiences by introducing tangibile programming toolkits~\cite{rodic_tangible_2022,papavlasopoulou_reviewing_2017,solid_beginning_2017}.
Regardless whether taught on its own or integrated in other subjects, teaching programming skills helps students develop important competences. In early childhood, the process of teaching programming benefits from enrichment with mixture with storytelling activities as a developmentally appropriate delivery approach. A systematic review from 2022 identified a need for the development of educational curricula integrating teaching of programming and the corresponding teacher-training programs~\cite{macrides_programming_2022}. In a discussion of the barriers to teaching programming to K-12 students, another meta-analysis draws attention to the highly abstract and complex syntax of text-based programming~\cite{sun_does_2023}. The authors indicate particularly the lack of student interest in text-based programming as one of the main challenges and suggest there is a need to improve teaching methods in order to foster student motivation and interest~\cite{sun_does_2023}.

Block-based and visual coding tools have emerged as a way to make programming more accessible to younger audiences that might struggle with the traditional text-based approach~\cite{hu_exploring_2021}. Research indicates that students in elementary schools who learned programming with block-based visual programming tools had better academic achievement than those who learned programming only with traditional text-based programming tools~\cite{hu_exploring_2021,hu_programming_2024}.
One of several toys following a similar rationale, Cubetto has been developed as an attempt to take this further~\cite{solid_beginning_2017}, making block-based programming accessible also to younger children who are especially susceptible to screen time overuse~\cite{wang_status_2023} and the surrounding adults that are increasingly reluctant to have young children engage with digital screens~\cite{gentleman_crux_2025}.
Documentation and learning materials around Cubetto in particular~\cite{solid_beginning_2017,solid_coding_2017} are rich with examples of different ways to engage with the robot when it comes to teaching programming skills to small children. This includes different physical activities for children, such as imitating and empathising with the robot, story co-creation activities, and creative activities - using drawing, role-play and others.

Research into the motivations to play games~\cite{yee_motivations_2005} suggests 3 broader groups of motivating categories: achievement, immersion and social motivation. These three categories and their 10 components give broader indication of what it takes to gamify a task. Certainly, challenge and achievement are part of these, and in our particular case these are addressed by the nature of navigating a robot. But the immersion component strongly features storytelling, containing topics such as discovery, role-play, customisation and escapism that could find form through narrative. Shared storytelling among the children could also boost the third category with the components of social motivation, containing the components socialising, relationship and teamwork. Finding ways to include any of these motivational factors reinforces the opportunity to engage children.
Our approach is inspired by the narrative success of AI Dungeon, an interactive platform that engages users with ChatGPT to collaboratively create narrative experiences~\cite{hua_playing_2020}.

\section{Method}

Here we aim to develop an early version of a formalised approach using rapid prototyping. To this end we engage in action research studying how to prompt LLMs, similar to the way this was previously done regarding the capabilities of a text-to-image transformer~\cite{ruskov_grimm_2023}. Such an approach allows us to simultaneously study both the potential of human-AI collaboration for preschool education using current LLMs, and the process of achieving this collaboration. By illustrating the use of five different models in parallel, we demonstrate that this process could be independent of the specific LLMs of choice.

\begin{figure}
  \centering
  \includegraphics[width=.65\linewidth,trim={0 3.5 0 3.5},clip]{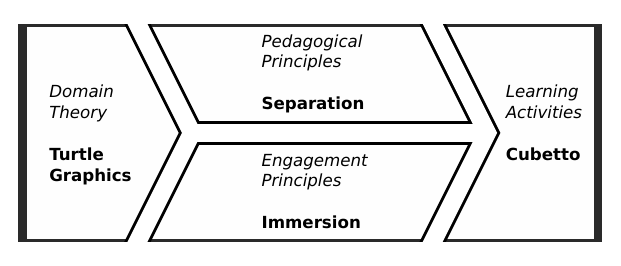}
  \caption{The pedagogical process integrating educational and game-based elements together.}
  \label{fig:model}
\end{figure}

To guide the development of the intended formalised approach, we turn to a methodological framework for the design of serious games that we have previously developed. We build a process proposed by Davies and Mangan that starts from theory and goes through pedagogical principles to finally arrive at the design of learning activities~\cite{davies_embedding_2008}, which we previously expanded by incorporating both pedagogical and engagement principles in parallel for the development of educational games~\cite{ruskov_employing_2014}. In particular, as pedagogical principles we previously proposed phenomenography, or variation theory, which postulates that in order to effectively understand a phenomenon, a learner needs to experience variation around it. Such variation encompasses three fundamental steps: \textit{(i) contrast} where one dimension of the phenomenon is played with, while others are maintained fixed; \textit{(iii) separation} where inversely one dimension is kept fixed; and \textit{(iii) fusion} where the free interplay between dimensions is experienced~\cite{marton_necessary_2014}. On the other hand, we also propose engagement principles, based on Yee's dimensions of achievement, immersion and social motivation~\cite{yee_motivations_2005}. Within such a framework, we aim to teach basic turtle programming with separation and immersion using Cubetto as illustrated in Figure~\ref{fig:model}.

\subsection{Materials}
\label{sec:materials}

\begin{figure}[b]
  \centering
  \includegraphics[width=.65\linewidth]{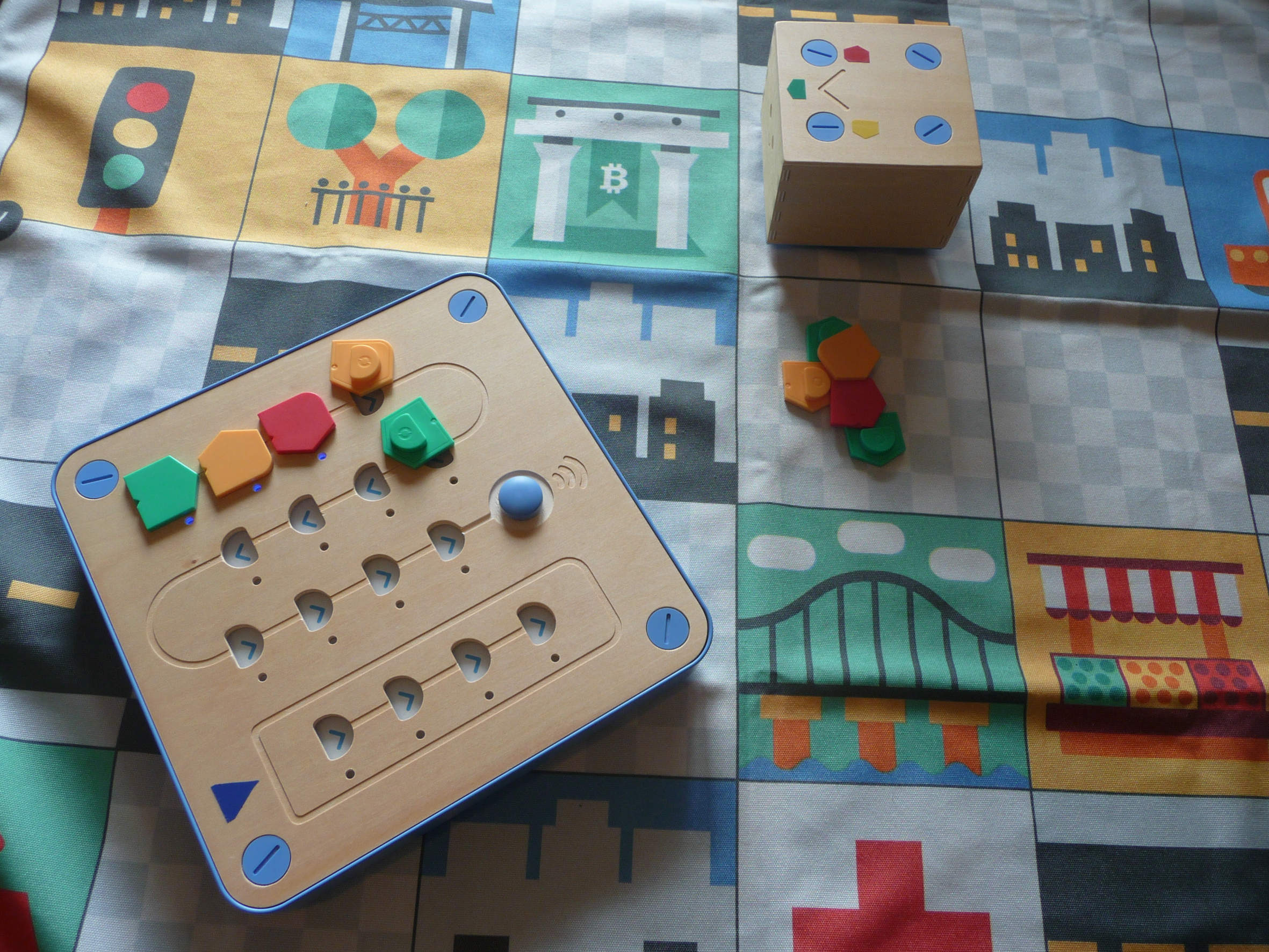}
  \caption{The enhanced Cubetto with color markers intended to help children match command bricks to actions.}
  \label{fig:cubetto}
\end{figure}

We propose a learning experience with Cubetto, a travelling box-shaped wooden robot. It follows instructions provided by placing physical command blocks on a control board. The standard Cubetto package comes with four commands, but for a preschool audience we use only the three that are about movement (\verb|move forward| - green, \verb|rotate left| - yellow and \verb|rotate right| - red). We also use a slightly enhanced version of Cubetto as visible in Figure~\ref{fig:cubetto}. In particular, to help children get used to the commands, we use colour markers on top the robot. These markers help associate the commands to the resulting actions.

As a way to address issues of sustained use and reproducibility~\cite{chen_how_2024}, for narrative generation our study involves exclusively open weights models. In particular, we consider Google's Gemma, Meta's Llama, Mistral, AllenAI's oLMo and Alibaba's Qwen. We access these using the llama.cpp platform\footnote{\url{https://llama-cpp-python.readthedocs.io}} and the GGUF model format. As a way to also ensure accessibility and data privacy, we focus on models that could be run locally on a contemporary laptop without using GPU functionality. These requirements take us to models within the 7-9 billion parameters range and their largest available 3-bit quantisation from HuggingFace\footnote{
\url{http://hf.co/bartowski/gemma-2-9b-it-GGUF}\\
\url{http://hf.co/bartowski/Meta-Llama-3.1-8B-Instruct-GGUF}\\
\url{http://hf.co/bartowski/Mistral-7B-Instruct-v0.3-GGUF}\\ 
\url{http://hf.co/bartowski/OLMo-2-1124-7B-Instruct-GGUF}\\
\url{http://hf.co/bartowski/Qwen2.5-7B-Instruct-GGUF}
}.  

Our experiments combine Cubetto with popular commonplace toys. On one hand these additional game subjects are meant as vehicles of inspirations and engagement, on the other as a way for children to overcome the challenge of programming being too abstract, particularly relevant for the age of interest for this study~\cite{rodic_tangible_2022,papavlasopoulou_reviewing_2017}. Furthermore, this gave us an opportunity to capture how well LLMs are able to handle practical constraints. To illustrate the approach, we choose four toys, popular among preschool children: Barbie dolls, Lego figures, Hot Wheels cars and Brio trains.
Each of these becomes a possible value for a parameter in our prompt as could be seen in the next subsection.
A particular challenge is presented by the \verb|Brio trains| subject. These wooden trains are challenging to adapt for two reasons: \textit{(i)} Cubetto can neither drive along, nor cross the bulky wooden tracks, and \textit{(ii)} trains follow tracks and thus do not have the affordance of directional movement represented by the \verb|rotate left| and \verb|rotate right| command blocks. Whereas this is intuitive to a person, it might not be to a LLM.

\subsection{Prompt template and parameters}

\begin{table*}
  \caption{Sample components of a story, provided as parameters to a prompt to propose a theme, a subject and a task to children.}
  \label{tab:params}
\begin{tabular}{llll}
\toprule
\textbf{\#} & \textbf{narrative world} & \textbf{subjects} & \textbf{task} \\
\midrule
\textbf{1} & knights and princesses & Barbie dolls & enact a pursuit \\
\textbf{2} & pirates & Lego figures & find a treasure \\
\textbf{3} & superheroes & Hot Wheels cars & enact a struggle \\
\textbf{4} & Wild West & Brio trains & rescue someone \\
\bottomrule
\end{tabular}
\end{table*}

For the composition of the prompt, we take inspiration from narrative structuralist traditions and more specifically the works of Vladimir Propp on morphology of folklore~\cite{propp_morphology_1968}. We devise a prompt template to have models generate ideas. In particular we adopt three personalisation parameters: \verb|narrative world|, \verb|subjects| and \verb|task|. The first two are used to set respectively the context and the protagonists so that they could possibly match the children's current preferences, aspirations and available physical toys. The third one, \emph{task}, serves to set the objective and possible suggestions for it are derived from Propp's \emph{functions}. We chose ones that we consider to relate to a protagonist, as illustrated in Table~\ref{tab:params}.

In our experimentation we use the combinations provided in the rows of Table~\ref{tab:params}, but the approach allows recombining them in any permutation or introducing further variants. We provide these parameters to LLMs through the following prompt:

``\emph{Suggest to a teacher a game with the Cubetto tangible programming toy for 
preschool children. It should be about \{narrative world\}, involve \{subjects\} and children should use 
Cubetto to \{task\}. Only three Cubetto command blocks should be used: forward, turn left and turn right. 
The description should be about half a page long.
}''

The exact reusable code for this procedure, as well as an overview of the resulting scenarios proposed by LLMs are available in appendix. This, and a full archive of the evolution of the prompts and corresponding generations are available at the online project repository\footnote{\url{https://github.com/mapto/our-coding-adventure}}.
\section{Results}

Our experiment involved seven rounds of generations to optimise the final results. This count includes only the end-to-end generation where the prompt was combined with all 4 sets of parameters from Table~\ref{tab:params}.

Generally all tested LLMs work well for this task and provide actionable instructions. Yet due to the issues outlined below, these should not be seen as proof-read and ready to use guides, and rather as creativity prompts for educators instead.

Without being prompted to do so, all models strive to produce structured, self-contained and detailed activity descriptions. These are formatted documents, having length of about one page. The structure of these produced documents is not consistent, even within iterations of the same LLMs. Yet, typical featured sections are scenario topic, objective(s), necessary materials, preparatory setup, instructions or gameplay, claimed learning outcomes, and variations of the proposed activity.

While this type of proposed solutions is generally useful, it leads to a technical problem - generated responses turn out to be larger than the available response token buffer of the LLMs so they get trimmed. On the upside, this typically happens at a point where the suggested task is sufficiently clear for a teacher to interpret despite the missing ending. We attempted to resolve the problem of too long responses by adding to the prompt, either requests for half-page answers, or impose a character limit. Gemma, Llama and oLMo appear to consistently ignore this part of the instructions and still produce responses longer than the buffer limit. Mistral and Qwen however, adapt their responses, providing shorter proposals and overcoming the problem.

Inconsistencies and hallucinations are widespread, yet straightforward to overcome by a teacher. In a commonly recurring example of an inconsistency, LLMs pretend to provide a list of required materials, but this is often incomplete, sometimes mentioning some necessary parts of the Cubetto package, but not all (e.g. interchangeably blocks and board, although one cannot function without the other). A typical hallucination is the proposed use of a Cubetto command block that does not exist. We addressed this by adding an explicit list of permissible command blocks, as detailed in Section~\ref{sec:materials}. This appears to have worked well and eliminated command-related hallucinations. However, considering that response length is an issue, LLMs consistently repeated the command-restricting requirement - something that we do not see as beneficial to the final task.

Task 1 (enact a pursuit) and task 3 (enact a struggle) could potentially present a challenge, as they could involve two active subjects, whereas models generally propose solutions involving a single Cubetto robot. All models, but Qwen transform task 1 into a rescue mission, i.e. the target does not move. Similarly, for task 3, only Mistral and Qwen do not explicitly transform the task into a rescue mission.
Coming back to the question of how many robots are used, the only ones that propose using more than one are oMLo for task 2 (find a treasure) and Qwen for task 3 (enact a struggle). These proposals involve dividing children in groups with a robot for each, even if these are not confronted in an interaction between groups. Thus, the proposed activities are easily adapted to using a single robot. More broadly speaking, all models but oLMo bring up the suggestion to divide children in teams. This makes unprovoked and unnecessary assumptions about the number of children involved and whether this number is sufficient and reasonable for a separation in teams.
When it comes to the particular challenge of the fourth task, Brio trains, Gemma suggests to ``Use Cubetto to guide a train through a Brio track'' which appears as difficult to follow through. Other suggestions might lead to the idea of using the Brio tracks as  continuous obstacles forming a canyon, which could be seen as a very creative solution. The other LLMs appear to be able to respond abstractly enough not to underline this particular challenge with providing phrases such as ``navigating through the Brio tracks'' (Llama), ``using Brio train tracks to create a winding path'' (Mistral).

\section{Discussion and Conclusion}

Our approach installs small models locally and we share all the materials necessary for the reproduction and adoption of this approach. However, should such a process be beyond the technical proficiency of teachers, online access to off-the-shelf models is a completely viable alternative using the same or similar prompts. The method does not disclose any personal information, so there should not be any privacy concerns.
However, a subsequent research phase needs to engage with a wider group of children and with pre-school teachers that are independent from the process developers. This would provide evidence about the usefulness of the approach in-the-wild and about perceptions of teachers as direct target audience, and of children as indirect audience. However, the presentation of this method needs to come in the context of the wider learning context of Cubetto, including other introductory activities to the robot~\cite{solid_beginning_2017,solid_coding_2017}.

We insist it is important to emphasise that so far there are no reasons or justifications in favour of exposing children directly to output generated by LLMs. Instead, as we propose here, the technology could be used to help overloaded teachers with rather trivial narratives that would put together elements that these same teachers consider useful to integrate in the learning process. Thus, we pursue a two-fold goal: on one hand to scaffold teacher creativity, on the other to not demand from them engagement with the activity that is beyond basic reasonable interaction.

In the process we have encountered cases where LLMs refuse to respond to tasks due to concerns about violence or discrimination. These were rare and circumvented by slight variations in the prompt. However, building on the aforementioned point of no direct exposure of children to LLM output, we believe the premise that censuring responses as a way to protect children is misplaced in educational contexts like ours. It is not up to the model or its authors to decide whether content is appropriate, nor muting conversations is an adequate way of addressing historical controversies. At least for the context in our consideration, we believe that it is teachers that need to take the lead on such decisions. This corresponds to previous research reporting that teachers need to be more involved in educational uses of LLMs~\cite{harvey_dont_2025}.

While definitely a topic of interest, due to the adopted method of this study, our results could not be used as a means of systematic comparison between LLMs. Yet, one cannot help but notice the emergence of consistently different behaviour by different models. Being different solutions to the same next token generation task, the differences emerge with the exact architectures and the exact training corpora and training procedures. While the research on optimal architectures is ongoing, there is less possibility to study how differences could be traced back to corpora or training procedures. Among the LLMs studied here, only the creators of oLMo are transparent about both the training process used and put an effort to make the training corpus publicly available~\cite{olmo_2_2025}. Nevertheless, this one example opens up to the theoretical possibility of studying how responses relate to specific instances in the training data.

\bibliography{references}

\begin{thebibliography}{22}
\expandafter\ifx\csname natexlab\endcsname\relax\def\natexlab#1{#1}\fi
\providecommand{\url}[1]{\texttt{#1}}
\providecommand{\href}[2]{#2}
\providecommand{\path}[1]{#1}
\providecommand{\DOIprefix}{doi:}
\providecommand{\ArXivprefix}{arXiv:}
\providecommand{\URLprefix}{URL: }
\providecommand{\Pubmedprefix}{pmid:}
\providecommand{\doi}[1]{\href{http://dx.doi.org/#1}{\path{#1}}}
\providecommand{\Pubmed}[1]{\href{pmid:#1}{\path{#1}}}
\providecommand{\bibinfo}[2]{#2}
\ifx\xfnm\relax \def\xfnm[#1]{\unskip,\space#1}\fi
\bibitem[{Harvey et~al.(2025)Harvey, Koenecke, and Kizilcec}]{harvey_dont_2025}
\bibinfo{author}{E.~Harvey}, \bibinfo{author}{A.~Koenecke}, \bibinfo{author}{R.~F. Kizilcec},
\newblock \bibinfo{title}{"{Don}'t {Forget} the {Teachers}": {Towards} an {Educator}-{Centered} {Understanding} of {Harms} from {Large} {Language} {Models} in {Education}},
\newblock in: \bibinfo{booktitle}{Proceedings of the 2025 {CHI} {Conference} on {Human} {Factors} in {Computing} {Systems}}, \bibinfo{publisher}{ACM}, \bibinfo{address}{Yokohama Japan}, \bibinfo{year}{2025}, pp. \bibinfo{pages}{1--19}. \DOIprefix\doi{10.1145/3706598.3713210}.
\bibitem[{Wang et~al.(2023)Wang, Qian, Li, and Wu}]{wang_status_2023}
\bibinfo{author}{C.~Wang}, \bibinfo{author}{H.~Qian}, \bibinfo{author}{H.~Li}, \bibinfo{author}{D.~Wu},
\newblock \bibinfo{title}{The status quo, contributors, consequences and models of digital overuse/problematic use in preschoolers: {A} scoping review},
\newblock \bibinfo{journal}{Frontiers in Psychology} \bibinfo{volume}{14} (\bibinfo{year}{2023}). \DOIprefix\doi{10.3389/fpsyg.2023.1049102}.
\bibitem[{Macrides et~al.(2022)Macrides, Miliou, and Angeli}]{macrides_programming_2022}
\bibinfo{author}{E.~Macrides}, \bibinfo{author}{O.~Miliou}, \bibinfo{author}{C.~Angeli},
\newblock \bibinfo{title}{Programming in early childhood education: {A} systematic review},
\newblock \bibinfo{journal}{Int. J. of Child-Computer Interaction} \bibinfo{volume}{32} (\bibinfo{year}{2022}). \DOIprefix\doi{10.1016/j.ijcci.2021.100396}.
\bibitem[{{Solid Labs}(2017{\natexlab{a}})}]{solid_beginning_2017}
\bibinfo{author}{{Solid Labs}}, \bibinfo{title}{Beginning computer programming for kids: An introductory guide to computational thinking and coding for kids aged 3-6 years old}, \bibinfo{publisher}{PRIMO Toys}, \bibinfo{year}{2017}{\natexlab{a}}. \URLprefix \url{https://www.primotoys.com/wp-content/uploads/2017/09/Ebook-PrimoToys_final-1.pdf}.
\bibitem[{{Solid Labs}(2017{\natexlab{b}})}]{solid_coding_2017}
\bibinfo{author}{{Solid Labs}}, \bibinfo{title}{Coding with Cubetto - Unit 1, Reception, Ages 4 to 5, UK National Curriculum}, \bibinfo{publisher}{PRIMO Toys}, \bibinfo{year}{2017}{\natexlab{b}}. \URLprefix \url{https://primotoys.com/education/resources/}.
\bibitem[{Hua and Raley(2020)}]{hua_playing_2020}
\bibinfo{author}{M.~Hua}, \bibinfo{author}{R.~Raley},
\newblock \bibinfo{title}{Playing {With} {Unicorns}: {AI} {Dungeon} and {Citizen} {NLP}},
\newblock \bibinfo{journal}{Digital Humanities Quarterly} \bibinfo{volume}{14} (\bibinfo{year}{2020}). \URLprefix \url{https://www.proquest.com/docview/2553526112}.
\bibitem[{Grover and Pea(2013)}]{grover_computational_2013}
\bibinfo{author}{S.~Grover}, \bibinfo{author}{R.~Pea},
\newblock \bibinfo{title}{Computational {Thinking} in {K}–12: {A} {Review} of the {State} of the {Field}},
\newblock \bibinfo{journal}{Educational Researcher} \bibinfo{volume}{42} (\bibinfo{year}{2013}) \bibinfo{pages}{38--43}. \DOIprefix\doi{10.3102/0013189X12463051}.
\bibitem[{McNerney(2004)}]{mcnerney_turtles_2004}
\bibinfo{author}{T.~S. McNerney},
\newblock \bibinfo{title}{From turtles to {Tangible} {Programming} {Bricks}: explorations in physical language design},
\newblock \bibinfo{journal}{Personal and Ubiquitous Computing} \bibinfo{volume}{8} (\bibinfo{year}{2004}) \bibinfo{pages}{326--337}. \DOIprefix\doi{10.1007/s00779-004-0295-6}.
\bibitem[{Rodić and Granić(2022)}]{rodic_tangible_2022}
\bibinfo{author}{L.~D. Rodić}, \bibinfo{author}{A.~Granić},
\newblock \bibinfo{title}{Tangible interfaces in early years’ education: a systematic review},
\newblock \bibinfo{journal}{Personal and Ubiquitous Computing} \bibinfo{volume}{26} (\bibinfo{year}{2022}) \bibinfo{pages}{39--77}. \DOIprefix\doi{10.1007/s00779-021-01556-x}.
\bibitem[{Papavlasopoulou et~al.(2017)Papavlasopoulou, Giannakos, and Jaccheri}]{papavlasopoulou_reviewing_2017}
\bibinfo{author}{S.~Papavlasopoulou}, \bibinfo{author}{M.~N. Giannakos}, \bibinfo{author}{L.~Jaccheri},
\newblock \bibinfo{title}{Reviewing the affordances of tangible programming languages: {Implications} for design and practice},
\newblock in: \bibinfo{booktitle}{2017 {IEEE} {Global} {Engineering} {Education} {Conference} ({EDUCON})}, \bibinfo{year}{2017}, pp. \bibinfo{pages}{1811--1816}. \DOIprefix\doi{10.1109/EDUCON.2017.7943096}.
\bibitem[{Sun and Zhou(2023)}]{sun_does_2023}
\bibinfo{author}{L.~Sun}, \bibinfo{author}{L.~Zhou},
\newblock \bibinfo{title}{Does text-based programming improve {K}-12 students’ {CT} skills? {Evidence} from a meta-analysis and synthesis of qualitative data in educational contexts},
\newblock \bibinfo{journal}{Thinking Skills and Creativity} \bibinfo{volume}{49} (\bibinfo{year}{2023}) \bibinfo{pages}{101340}. \DOIprefix\doi{10.1016/j.tsc.2023.101340}.
\bibitem[{Hu et~al.(2021)Hu, Chen, and Su}]{hu_exploring_2021}
\bibinfo{author}{Y.~Hu}, \bibinfo{author}{C.-H. Chen}, \bibinfo{author}{C.-Y. Su},
\newblock \bibinfo{title}{Exploring the {Effectiveness} and {Moderators} of {Block}-{Based} {Visual} {Programming} on {Student} {Learning}: {A} {Meta}-{Analysis}},
\newblock \bibinfo{journal}{Journal of Educational Computing Research} \bibinfo{volume}{58} (\bibinfo{year}{2021}) \bibinfo{pages}{1467--1493}. \DOIprefix\doi{10.1177/0735633120945935}.
\bibitem[{Hu(2024)}]{hu_programming_2024}
\bibinfo{author}{L.~Hu},
\newblock \bibinfo{title}{Programming and 21st century skill development in {K}‐12 schools: {A} multidimensional meta‐analysis},
\newblock \bibinfo{journal}{J. of Computer Assisted Learning} \bibinfo{volume}{40} (\bibinfo{year}{2024}) \bibinfo{pages}{610--636}. \DOIprefix\doi{10.1111/jcal.12904}.
\bibitem[{Gentleman(2025)}]{gentleman_crux_2025}
\bibinfo{author}{A.~Gentleman},
\newblock \bibinfo{title}{‘the crux of all evil’: what happened to the first city that tried to ban smartphones for under-14s?},
\newblock \bibinfo{journal}{The Guardian}  (\bibinfo{year}{2025}). \URLprefix \url{https://www.theguardian.com/technology/2025/may/07/the-crux-of-all-evil-what-happened-to-the-first-city-that-tried-to-ban-smartphones-for-under-14s}.
\bibitem[{Yee(2005)}]{yee_motivations_2005}
\bibinfo{author}{N.~Yee},
\newblock \bibinfo{title}{Motivations of {Play} in {MMORPGs}},
\newblock in: \bibinfo{booktitle}{{DiGRA} 2005 {Conference}}, \bibinfo{year}{2005}, p.~\bibinfo{pages}{46}. \URLprefix \url{http://www.nickyee.com/daedalus/motivations.pdf}.
\bibitem[{Ruskov(2023)}]{ruskov_grimm_2023}
\bibinfo{author}{M.~Ruskov},
\newblock \bibinfo{title}{Grimm in {Wonderland}: {Prompt} {Engineering} with {Midjourney} to {Illustrate} {Fairytales}},
\newblock in: \bibinfo{editor}{B.~Alessia}, \bibinfo{editor}{F.~Alex}, \bibinfo{editor}{F.~Stefano}, \bibinfo{editor}{M.~Stefano}, \bibinfo{editor}{R.~Domenico} (Eds.), \bibinfo{booktitle}{Proceedings of the 19th {Conference} on {Information} and {Research} {Science} {Connecting} to {Digital} and {Library} {Science}}, volume \bibinfo{volume}{3365} of \textit{\bibinfo{series}{{CEUR} {Workshop} {Proceedings}}}, \bibinfo{year}{2023}, pp. \bibinfo{pages}{180--191}. \URLprefix \url{https://ceur-ws.org/Vol-3365/#paper6}.
\bibitem[{Davies and Mangan(2008)}]{davies_embedding_2008}
\bibinfo{author}{P.~I. Davies}, \bibinfo{author}{J.~Mangan},
\newblock \bibinfo{title}{Embedding {Threshold} {Concepts}: from theory to pedagogical principles to learning activities},
\newblock in: \bibinfo{editor}{R.~Land}, \bibinfo{editor}{J.~H. Meyer}, \bibinfo{editor}{J.~Smith} (Eds.), \bibinfo{booktitle}{Threshold Concepts within the Disciplines}, \bibinfo{publisher}{Brill}, \bibinfo{address}{Rotterdam}, \bibinfo{year}{2008}, pp. \bibinfo{pages}{37--50}. \DOIprefix\doi{10.1163/9789460911477_004}.
\bibitem[{Ruskov(2014)}]{ruskov_employing_2014}
\bibinfo{author}{M.~Ruskov}, \bibinfo{title}{Employing {Variation} in the {Object} of {Learning} for the {Design}-based {Development} of {Serious} {Games} that {Support} {Learning} of {Conditional} {Knowledge}}, \bibinfo{type}{{PhD} {Thesis}}, University College London, \bibinfo{year}{2014}. \URLprefix \url{https://discovery.ucl.ac.uk/id/eprint/1457529/}.
\bibitem[{Marton(2014)}]{marton_necessary_2014}
\bibinfo{author}{F.~Marton}, \bibinfo{title}{Necessary {Conditions} of {Learning}}, \bibinfo{publisher}{Taylor and Francis}, \bibinfo{address}{Hoboken}, \bibinfo{year}{2014}.
\bibitem[{Chen et~al.(2024)Chen, Zaharia, and Zou}]{chen_how_2024}
\bibinfo{author}{L.~Chen}, \bibinfo{author}{M.~Zaharia}, \bibinfo{author}{J.~Zou},
\newblock \bibinfo{title}{How {Is} {ChatGPT}’s {Behavior} {Changing} {Over} {Time}?},
\newblock \bibinfo{journal}{Harvard Data Science Review} \bibinfo{volume}{6} (\bibinfo{year}{2024}). \DOIprefix\doi{10.1162/99608f92.5317da47}.
\bibitem[{Propp(1968)}]{propp_morphology_1968}
\bibinfo{author}{V.~Propp}, \bibinfo{title}{Morphology of the {Folktale}}, \bibinfo{edition}{2} ed., \bibinfo{publisher}{University of Texas Press}, \bibinfo{year}{1968}.
\bibitem[{Walsh et~al.(2025)Walsh, Soldaini, Groeneveld, Lo, Arora, Bhagia, Gu, Huang, Jordan, Lambert et~al.}]{olmo_2_2025}
\bibinfo{author}{P.~Walsh}, \bibinfo{author}{L.~Soldaini}, \bibinfo{author}{D.~Groeneveld}, \bibinfo{author}{K.~Lo}, \bibinfo{author}{S.~Arora}, \bibinfo{author}{A.~Bhagia}, \bibinfo{author}{Y.~Gu}, \bibinfo{author}{S.~Huang}, \bibinfo{author}{M.~Jordan}, \bibinfo{author}{N.~Lambert}, et~al., \bibinfo{title}{2 {OLMo} 2 {Furious}}, \bibinfo{year}{2025}. \DOIprefix\doi{10.48550/arXiv.2501.00656}.

\end{thebibliography}


\appendix
\pagebreak

\section{Implementation}
The locally executable code that was used for the analysis is provided below.

\subsection{Requirements}

\begin{verbatim}
llama-cpp-python==0.3.9
huggingface-hub==0.31.4
\end{verbatim}

\subsection{Script}

\begin{lstlisting}[basicstyle=\ttfamily\scriptsize,language=Python]
from llama_cpp import Llama


models = [
    {"repo": "bartowski/gemma-2-9b-it-GGUF",              "file": "gemma-2-9b-it-Q3_K_XL.gguf"},
    {"repo": "bartowski/OLMo-2-1124-7B-Instruct-GGUF",    "file": "OLMo-2-1124-7B-Instruct-Q3_K_XL.gguf"},
    {"repo": "bartowski/Mistral-7B-Instruct-v0.3-GGUF",   "file": "Mistral-7B-Instruct-v0.3-Q3_K_L.gguf"},
    {"repo": "bartowski/Qwen2.5-7B-Instruct-GGUF",        "file": "Qwen2.5-7B-Instruct-Q3_K_XL.gguf"},
    {"repo": "bartowski/Meta-Llama-3.1-8B-Instruct-GGUF", "file": "Meta-Llama-3.1-8B-Instruct-Q3_K_XL.gguf"},
]

params = [
    {"world": "pirates",                "objects": "Lego figures",    "task": "find a treasure"},
    {"world": "Wild West",              "objects": "Brio trains",     "task": "rescue someone"},
    {"world": "superheroes",            "objects": "Hot Wheels cars", "task": "enact a struggle"},
    {"world": "knights and princesses", "objects": "Barbie dolls",    "task": "enact a pursuit"},
]

prompt_templ = """Provide a description of a game with the Cubetto tangible programming toy for preschool children. It should be about {world}, involve {objects} and children should use Cubetto to {task}. Only three Cubetto command blocks should be used: forward, turn left and turn right."""


for m in models:
    mname = m["repo"].split("/")[1].split("-")[0]
    print(mname)
    
    llm = Llama.from_pretrained(
        repo_id=m["repo"],
        filename=m["file"],
        n_gpu_layers=-1,
        flash_attn=True,
    )
    
    for p in params:
        print(p)
        prompt = prompt_templ.format(**p)
        msg = [{"role": "user", "content": prompt}]
        
        output = llm.create_chat_completion(messages=msg, temperature=0)
        result = output["choices"][0]["message"]["content"]
        
        fname = f'results/{mname}-{p["world"]}-{p["objects"]}-{p["task"]}.txt'
        with open(fname, "w") as fout:
            fout.write(f"{m['repo'].split('/')[1]}\n{prompt}\n\n-------------------\n{result}")
\end{lstlisting}

\section{Outputs}

On the next page, the final produced output per model and per task, the last two being indicated in the first paragraph of each minipage. This seen in better detail, and documentation of previous iterations is available at the project repository\footnote{\url{https://github.com/mapto/our-coding-adventure}}.





\noindent
\begin{figure}[b]
  \centering
  \includegraphics[width=1.15\textwidth,trim={0 60 0 19},clip]{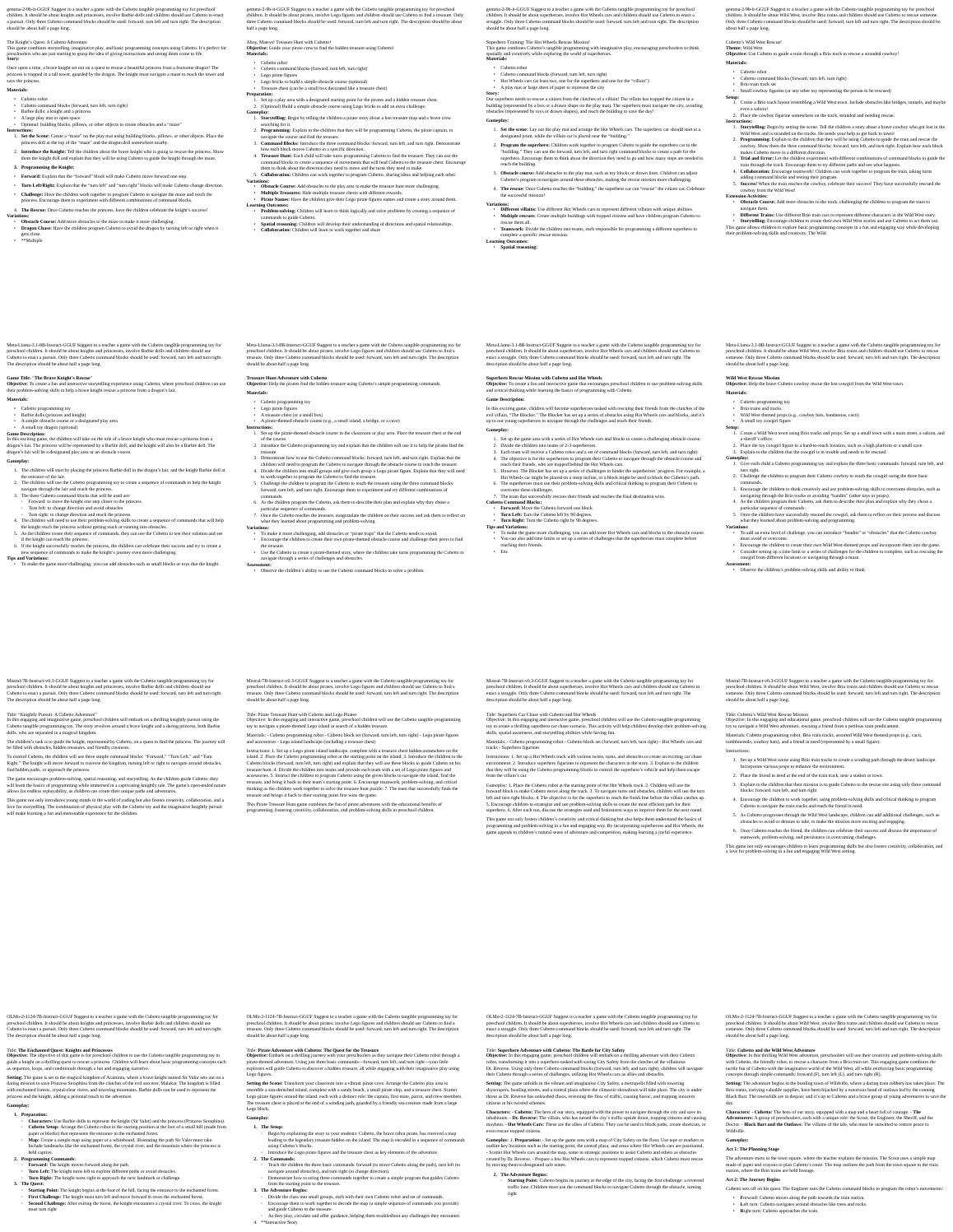}
  \includegraphics[width=1.15\linewidth,trim={0 690 0 19},clip]{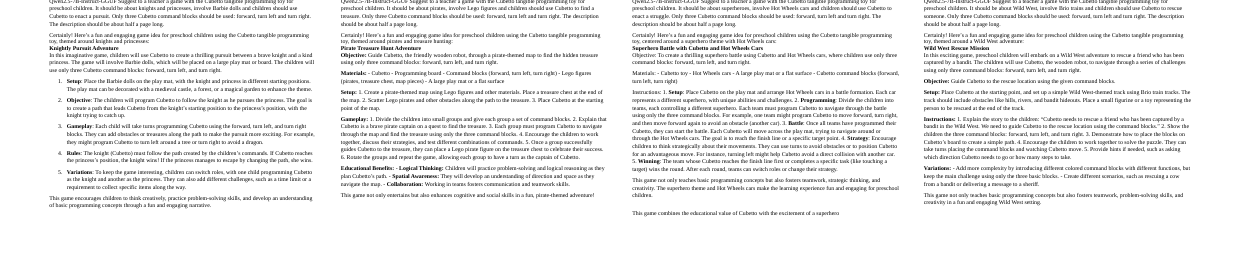}
  \label{fig:output}
\end{figure}


\end{document}